# Response to Comment on "Combined open shell Hartree-Fock theory of atomic-molecular and nuclear systems" [I.I.Guseinov, J. Math. Chem., 42 (2007) 177] by B. N. Plakhutin and E. R. Davidson

I. I. Guseinov

*Department of Physics, Faculty of Arts and Sciences, Onsekiz Mart University, Çanakkale, Turkey*

**Abstract**

This article is a thorough critique to the Plakhutin-Davidson's comments made to our paper published in the recent year. A detailed critical examination of the arguments that led to the suggested comments by Plakhutin and Davidson reveals some serious flaws. It is demonstrated that *the principle of the indistinguishability of identical particles* is not taking into account in Roothaan's open shell theory. This principle leads to the fact that the orbital-dependent energy functional and, therefore, the Hartree-Fock and Hartree-Fock-Roothaan equations for open shell systems presented by Roothaan and others are not, in general, invariant under unitary transformation of the combined closed-open shells orbitals. From a mathematical point of view this statement is fundamentally flawless. It is shown that the Plakhutin-Davidson's personal views about our assumptions concerning the insufficiencies of classic Roothaan's open-shell theory are *undisputedly* wrong.

## 1. Introduction

The Hartree-Fock (HF) and Hartree-Fock-Roothaan (HFR) equations for a ground state must be derived by requiring that the identical particles of a system obey the principle of indistinguishability which plays a fundamental part of Quantum Mechanics [1]. In the case of closed shell systems, these theories have been constructed on the basis of this principle.

It is well-known that the Roothaan's open shells treatment is an extension of HF theory for closed shells systems. However, more complex problems arise in Roothaan's extension and in the extensions to arbitrary open shell states by others because the principle of indistinguishability of identical particles is not included in these theories. These extensions cannot be used especially in the case of *nonsymmetrical coupling coefficients* which must also include so called "*non-Roothaan*" states. All of the previous well-accepted results are obtained only in the special case of open shells having the symmetrical coupling coefficients which satify the principle of indistinguishability. The authors of Comment paper [3] try to demonstrate that the expressions for the total energy functional and HF equations which contain the nonsymmetrical coupling coefficients are, *in sharp contrast to the principle of*

*indistinguishabilty of similar particles*, invariant under unitary transformations of the orbitals. Based on these *undisputedly wrong* assumptions, Plakhutin and Davidson are trying to destroy our recent studies on combined open shell HF and HFR (CHF and CHFR) theory of atomic-molecular and nuclear systems [4, 5] based on the use of principle of indistinguishabilty of similar particles. In previous papers [6-11], the CHFR method was applied and tested successfully for some arbitrary open shells atoms and molecules.

In next section, by taking into account the principle of indistinguishabilty of identical particles of Quantum Mechanics, we demonstrate all of the insufficiencies arising in open-shell SCF theory first treated by Roothaan in his classic paper and in its present-day formulation. Thus, the Plakhutin-Davidson's personal assumptions concerning our "Combined open shell Hartree-Fock theory of atomic-molecular and nuclear systems" are *undisputedly wrong*.

## 2. Source of insufficiencies of Roothaan's open shell HF theory

Let us try to show that the Roothaan's open shell theory does not satisfy the principle of indistinguishabilty of similar particles. To do so, it is natural to start from the energy functional suggested by Roothaan in Ref. [2] and its present-day formulation (see Eq. (1) in [3]), respectively,

$$E = 2\sum_{k} H_k + \sum_{kl}(2J_{kl} - K_{kl}) + f\left[2\sum_{m} H_m + f\sum_{mn}(2aJ_{mn} - bK_{mn}) + 2\sum_{km}(2J_{km} - K_{km})\right] \quad (1)$$

$$E = 2\sum_{k} f_k H_k + \sum_{kl} f_k f_l (2a_{kl} J_{kl} - b_{kl} K_{kl}). \quad (2)$$

The indices ($k,l$) and ($m,n$) in Eq.(1) numerate the occupied closed-shell and open-shell orbitals, respectively. The coupling coefficients $a_{kl}$ and $b_{kl}$ which must also include so called "*non-Roothaan*" states are not, in general, symmetrical with respect to the indices $k$ and $l$, i.e., $a_{kl} \neq a_{lk}$ and $b_{kl} \neq b_{lk}$.

It is easy to see that the energy functionals defined by Eqs. (1) and (2) are not invariant with respect to the interchange of the states and, therefore, also are not invariant under unitary transformations of ensemble sets for closed and open shells orbitals. Hence, the application of the variational principle to the energy functionals of the Roothaan's theory gives the two kinds of Fock matrices (a closed and an open shell one) depending on the implementation, combined together to obtain a unique set of orbitals and eigenvalues (see Eqs.(25a) and (25b) in [2]. Thus, the energy functionals defined by Eqs. (1) and (2), therefore, the Fock operators obtained from these functionals are not invariant with respect to the interchange of states and the unitary transformations of components in the ensemble of closed and open shells orbitals. Clearly, a formulation of the HF theory in which the closed- and open-shell orbitals are the eigenfunctions of the same Fock operator would be desirable. In our papers [4, 5], by the use of principle of indistinguishabilty of identical particle we eliminated all of the insufficiencies arising in Roothaan's open shell HF theory by introducing the *new energy functional*, and derived the combined open shell HF and HFR equations for atomic-molecular and nuclear systems applicable to any multideterminantal state of a single configuration that has any number of open shells of any symmetry.